\documentclass[11pt]{article}

\usepackage{amssymb,amsmath}                                           
\usepackage{bbm}                                                       
\usepackage[usenames,dvipsnames]{color}                                
\usepackage{setspace}                                                  
\usepackage{hyperref}                                                  
\usepackage[left= 1in,right=1in,top=1in,bottom=1in]{geometry}          
\usepackage{comment}                                                   
\usepackage{authblk}                                                   
\usepackage{graphicx}                                                  
\usepackage{caption}                                                    
\usepackage{subcaption}
\usepackage{enumerate}													
\usepackage{epigraph}
\usepackage[utf8]{inputenc}
\usepackage[english]{babel}
\usepackage{csquotes}
\usepackage{natbib}

\hypersetup{                 
    colorlinks=true,         
    linkcolor=blue,          
    citecolor=red,           
    urlcolor=Violet          
}

\let\oldmarginpar\marginpar
\renewcommand\marginpar[1]{\oldmarginpar{\color{red}\raggedright\scriptsize #1}}
\newcommand{\pb}[2]{\ensuremath{\lf\{#1,#2 \rt\}}}

\def\lf {\ensuremath{\left}}											
\def\rt {\ensuremath{\right}}
\def\de {{\rm d}}														

\title{New Difficulties for the Past Hypothesis}

\author[1,2]{{\bf Sean Gryb}\thanks{email: \href{mailto:sean.gry@gmail.com}{sean.gryb@gmail.com}, website: \href{https://seangryb.wordpress.com}{seangryb.wordpress.com}}}
\affil[1]{{\it Faculty of Philosophy}, University of Groningen}
\affil[2]{ {\it Van Swinderen Institute for Particle Physics and Gravity}, University of Groningen }

\date{}

\begin{document}

\maketitle

\begin{abstract}
	Many macroscopic physical processes are known to occur in a time-directed way despite the apparent time-symmetry of the known fundamental laws. A popular explanation is to postulate an unimaginably atypical state for the early universe --- a `Past Hypothesis' (PH) --- that seeds the time-asymmetry from which all others follow. I will argue that such a PH faces serious new difficulties. First I strengthen the grounds for existing criticism by providing a systematic analytic framework for assessing the status of the PH. I outline three broad categories of criticism that put into question a list of essential requirements of the proposal. The resulting analysis paints a grim picture for the prospects of providing an adequate formulation for an explicit PH. I then provide a new argument that substantively extends this criticism by showing that any time-independent measure on the space of models of the universe must necessarily break one of its gauge symmetries. The PH then faces a new dilemma: reject a gauge symmetry of the universe and introduce a distinction without difference or reject the time-independence of the measure and lose explanatory power.
\end{abstract}

\tableofcontents
\clearpage

\section{Introduction} 
\label{sec:intro}

Everyday processes occur in such a way as to suggest an obvious intuitive difference between the past and the future. One of the great mysteries of physics and, in particular, the metaphysics of time is to explain the existence of this time-asymmetry despite the symmetry of the known microscopic theories of physics under an appropriate time reversal operation. Ludwig Boltzmann provided a proposal for such an explanation that seems to work for everyday processes.\footnote{See \cite{sep-statphys-Boltzmann}.} This proposal placed the burden of explanation not on the nature of the fundamental laws but on the nature of the initial state. The time-symmetry of the laws is then broken by the asymmetrical restriction to possible models that have highly atypical initial (but not final) states. In this way, Boltzmann attempted to explain why one might readily expect a cup of coffee to fall and shatter onto the ground but would not expect a mess of coffee and shards of cup to reassemble themselves. Because the cup of coffee is a highly unusual state in the space of possible ways that the constituents of the cup and coffee could be arranged, it is more typical to see the pieces scatter haphazardly than to see then reassemble as a cup of coffee.

While this kind of explanation works reasonably well for simple thermodynamic systems, complications arise when attempting to apply this strategy to the universe as a whole. Evidence from modern cosmology that the earliest known states of the universe appear to have extremely low entropy seems to have improved the situation. Positing an unimaginably atypical past state for the entire universe, a so-called \emph{Past Hypothesis (PH)} \citep{albert2009time}, might then be used to iteratively provide an explanation for why nested subsystems of the universe --- such as a coffee cup in a room in a city on a planet etc --- should individually be expected to start off in atypical states. Early versions of the PH date back to Boltzmann himself \citeyearpar{boltzmann2012suicide} and comprehensive improvements making use of modern lessons from cosmology have been advanced mostly notably by Roger Penrose \citeyearpar{Penrose:1979WCH,penrose1994second}, Joel Lebowitz \citeyearpar{lebowitz1993boltzmann}, Shelly Goldstein \citeyearpar{goldstein2001boltzmann,goldstein2004boltzmann} and Huw Price \citeyearpar{price1997book,price2002boltzmann,price2004origins}. A well-known formulation has been advocated in \citet{albert2009time} where the phrase `Past Hypothesis' was coined after an initial proposal by Richard Feynman \citeyearpar[p.116]{feynman2017character}.

The status of the PH remains controversial: it is not difficult to find both glowing appraisals and scathing criticism. Barry Loewer rates the problem of time-asymmetry as ``among the most important questions in the metaphysics of science'' \citep{loewer2012two} and the PH as ``the most promising approach to reductive accounts of time's arrows''. Huw Price rates the discovery of the low entropy past ``one of the most important [achievements] in the entire history of physics''\citeyearpar{price2004origins}. Despite these grand claims, criticism abounds. John Earman \citeyearpar{earman2006past} puts it bluntly:
\begin{quote}
	This dogma, I contend, is ill-motivated and ill-defined, and its implementation consists mainly in furious hand waving and wishful thinking. In short, it is (to borrow a phrase from Pauli) not even false.
\end{quote}
\cite{Wald:2012zf} deliver a scathing critique of the basic technical premises of the idea identifying ``a number of serious difficulties in'' attempting to formulate concrete implementations of the proposal.

The purpose of this paper is to asses and extend existing criticism and introduce a particularly troubling dilemma in order to argue that the PH faces disturbing new difficulties. First we will provide a comprehensive analysis of existing criticism of the PH for the purpose of assessing its status. Three broad categories of criticism are identified and listed at the beginning of \S\ref{sec:deconstructing_the_argument}. These categories provide a formal scheme for describing and evaluating different criticisms of the PH that have been advanced in the literature. To add precision to this process, we will start in \S\ref{sec:the_past_hypothesis} by giving a modern presentation of the arguments motivating the PH and identify a list of important conditions (in \S\ref{sub:key_assumptions_of_the_past_hypothesis}) that underly these arguments. We will then analyze several examples of criticism, taken as exemplars, in each category by identifying the specific conditions that each criticism puts into question. While this list of criticisms is not meant to be exhaustive and no single form of criticism should be seen as providing grounds to reject the entire proposal, when taken together these objections are sufficient to raise serious concerns regarding the PH. The resulting analysis already paints a rather grim picture for the prospects of formulating a PH in an unambiguous way using sound mathematical and physical principles.

One common response to such objections is that they amount merely to an unreasonable insistence on technical rigour given the immense mathematical difficulties associated with defining measures in general relativity. In response to such objections, we show in \S\ref{sec:symmetries_and_measure_ambiguities} that the PH encounters a troubling dilemma that persists even if all such technical concerns are removed. This dilemma is an uncomfortable choice between a loss of explanatory power --- the \emph{first horn} (see \S\ref{sub:the_origin_of_measure_ambiguities_in_cosmology}) --- and the breaking of a gauge symmetry --- the \emph{second horn} (see \S\ref{sub:symmetry_and_ambiguity}).

To establish this dilemma, we begin by using the analysis of \S\ref{sec:the_past_hypothesis} and \ref{sec:deconstructing_the_argument} to describe the first horn. In \S\ref{sec:the_past_hypothesis} we show that it is essential to the arguments of the PH to provide a justification for the measure used in the required typicality argument. Then in \S\ref{sec:deconstructing_the_argument} and \S\ref{sub:the_origin_of_measure_ambiguities_in_cosmology} we argue that the existence of a unique time-independent measure on the cosmological state space is essential to the explanatory claims of the PH. In \S\ref{sub:dynamical_similarity_in_the_universe} we show that the unique time-independent measure is not invariant under a particular cosmological symmetry called \emph{dynamical similarity}. Using this, we establish the second horn of the dilemma in \S\ref{sub:symmetry_and_ambiguity} by arguing that a failure of the measure to be invariant under this symmetry introduces a distinction without difference by over-counting empirically indistinguishable states. This leads to the following dilemma: either reject a time-independent measure and undermine the explanatory basis for the PH (horn 1) or introduce a distinction without difference by breaking dynamical similarity (horn 2).

\section{The Past Hypothesis} 
\label{sec:the_past_hypothesis}


In this section we will first provide a modern outline of Boltzmann-style explanations of time-asymmetry, \S\ref{sec:preliminaries}, and then use this framework to illustrate the basic logic of the Past Hypothesis, \S\ref{sub:the_past_hypothesis}. We compile a list (\S\ref{sub:key_assumptions_of_the_past_hypothesis}) of conditions necessary for the arguments of the PH collected from \S\ref{sub:the_past_hypothesis}.

\subsection{Boltzmannian explanations of time-asymmetry}
\label{sec:preliminaries}

In the Boltzmannian reasoning, the ultimate goal is to explain within a given system the time-asymmetry of some macroscopic processes from the fundamentally time-symmetric microscopic processes that underly it. The main formal ingredients of this procedure therefore involve a specification of the macro- and micro-states of the system, a particular reductive map between them, and a way to describe their behavior. This is usually achieved in the context of the Hamiltonian formalism. In this formalism, the micro-states of the systems in question are given in terms of representations of the configurations of the microscopic constituents of the system and their states of motion. These are expressed as generalized position and momentum variables formally represented by a symplectic manifold, $\Gamma$, that specifies the \emph{phase space} of the system. A phase space of this kind has a number of interesting mathematical properties. Of central importance is the existence of a privileged measure, called the \emph{Liouville measure} $\mu_L(\Sigma)$, that can be used to assign weights to arbitrary regions $\Sigma \in \Gamma$. The Liouville measure is singled-out by its rather remarkable symmetry properties that will be discussed in detail below. Concretely, the Liouville measure is the integral over the $n^\text{th}$ power of the symplectic form, where $n$ is half the dimension of $\Gamma$. In \emph{Darboux} coordinates $(q_i,p_i)$ where $\pb {q_i}{p_j} = \delta_{ij}$, we have $\mu_L(\Sigma) = \int_\Sigma \prod_{i = 1}^n \de p_i\, \de q_i $ (i.e., $\mu_L$ is the Lebesgue measure on $\Gamma$ in these coordinates). For systems with infinite degrees of freedom or where the range of positions and momenta is infinite, there may be mathematical difficulties in precisely formulating this measure. The first set of relevant conditions for applying the Boltzmannian logic is therefore that there exists some way of writing a mathematically precise (Condition-\ref{assumption:rigor}) and empirically unambiguous (Condition-\ref{assumption:uniqueness}) measure $\mu$ on $\Gamma$. (Note that this does not necessarily have to be the Liouville measure.)

With a suitable measure in hand one can assign weights to arbitrary regions in phase space. These weights can be taken to define different notions of \emph{typicality} for these regions. For example, one can say that a particular region $A$ is \emph{typical} on phase space if its weight as determined by $\mu$ is sufficiently large with respect to the weight of phase space itself:
\begin{equation}
	\frac{\mu(\Gamma) - \mu(A)}{\mu(\Gamma)} \ll 1\,.
\end{equation}
In general, a set $S$ is said to be typical with respect to some property $P$ and measure $\mu$ if its weight according to $\mu$ is large as compared with all other sets that possess the property $P$ \citep{frigg2009typicality}. Clearly, any notion of typicality requires some interpretation for the weights provided by $\mu$ in order to have any meaning. For the purposes of Boltzmann's argument, we will see below that it will be necessary to interpret the weight $\mu(\Sigma)$ as the relative likelihood of finding the system in a particular region $\Sigma$ (as opposed to somewhere else in $\Gamma$) at any given time. We identify this as an additional requirement (Condition-\ref{assumption:Boltzmann 2}) of the formalism.

The next formal step is to define the macro-states of a system. Physically these correspond to macroscopic states of the system such as temperature, volume, pressure, etc. Formally they are represented by some macro-state space $M$ which must have a (much) smaller dimension than $\Gamma$. Because Boltzmann was usually considering closed systems where the total energy $E$ is preserved, it is customary to consider states restricted to constant energy surfaces $\Gamma_E = \Gamma|_{E = \text{constant}}$ (i.e., the micro-canonical ensemble). In general many microscopic states will be indistinguishable from each other at the macroscopic level. This indistinguishability is modeled as a projection from $\Gamma_E$ to $M$. The micro-states identified under this projection define a partitioning of $\Gamma_E$ into the partitions $\Gamma_m$, where $m \in M$ ranges over all macro-states in $M$. These partitions represent equivalence classes of macroscopically indistinguishable micro-states. In order for these to be meaningful physically, there must exist some epistemologically motivated coarse-graining procedure that realizes this projection. For example, if the macroscopic variable in question is the temperature, then the temperature must be a well-defined quantity. We identify this requirement with a further condition (Condition-\ref{assumption:epistomology}). With these ingredients in hand it is now possible to define the \emph{Boltzmann entropy} (from now on called the `entropy' unless otherwise stated) of a particular macro-state $m$ as the logarithm of the Liouville weight of the partition $\Gamma_m$:\footnote{$k_B$ fixes the units of $S_\text{B}$.}
\begin{equation}\label{eq:entropy def}
 	S_\text{B} = k_\text{B} \log[ \mu_L(\Gamma_{m}) ]\,.
\end{equation} 

\begin{figure}
	\begin{center}
		\includegraphics[width=0.618\textwidth]{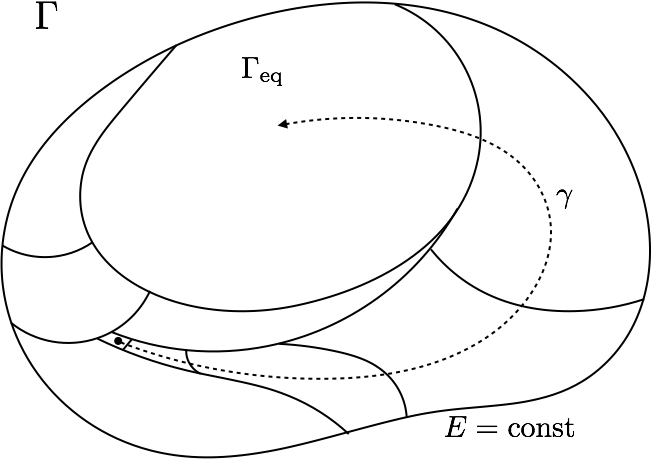}
		\caption{\label{fig:phase space} A small, atypical initial state will typically spend most of its future in a large equilibrium state $\Gamma_\text{eq}$.}
	\end{center}
\end{figure}

The last formal ingredient describes the behavior of the system. Consider representing a single history of the system by a curve $\gamma$ in $\Gamma$ as in Fig~\ref{fig:phase space}. The dynamics of an entire region can then be understood in terms of a collection of curves or a \emph{flow} where each point in the region is mapped to another neighboring point on $\Gamma$. For systems where the energy is conserved, this flow can be expressed mathematically in terms of a single phase space function, $H$, called the \emph{Hamiltonian} of the system. A theorem of primary importance due to \cite{liouville1838note} shows that the flow generated by \emph{any} choice of Hamiltonian function is guaranteed to preserve the Liouville measure. An immediate caveat of this theorem is that, up to a constant, the Liouville measure is the unique (smooth) measure preserved by any choice of Hamiltonian.\footnote{Proof: Formally Liouville's theorem implies $\mathcal L_{\chi_H} \mu = 0\,,\forall H: \Gamma \mapsto \mathbbm R$ where $\mu = \omega^n$ and $\omega$ is the symplectic 2-form on $\Gamma$ and the vector field $\chi_H$ is determined via $\de H = \iota_{\chi_H} \omega$. Writing an arbitrary smooth volume-form as $v = f \mu$, where $f$ is some arbitrary smooth positive function $f:\Gamma \mapsto \mathbbm R^+$, then Liouville's theorem and the condition $\mathcal L_{\chi_H} v = 0$ immediately lead to $f =$ constant.} It is this property that mathematically privileges the Liouville measure. Liouville's theorem is therefore doubly important for Boltzmann's reasoning. It provides at the same time a possible justification, via uniqueness, for the choice of typicality measure $\mu$ and a consistency requirement, via the invariance property under evolution, for being able to use the same measure at different times. The latter point arises as a consequence of a stronger requirement, which we identify as Condition-\ref{assumption:invariance}, that the typicality measure be invariant under all gauge symmetries of the system (in this case time-translational and, crucially, time-reversal invariance). In this context and for the remainder of the paper, we will understand a `gauge-symmetry' to be a transformation of the representations of a system that relates physically indistinguishable states.

We are now equipped to give a modern synthesis of Boltzmann's reasoning. First one must show that for the system in question there exists an exceptionally large macro-state $\Gamma_\text{eq}$ that takes up most of the phase space volume of the system. We take this to represent a further requirement that $\Gamma_\text{eq}$ be a typical state in $\Gamma_E$ (Condition-\ref{assumption:Boltzmann 1}). The relevance of Condition-\ref{assumption:Boltzmann 1} can be seen by the interpretation given to the weights of $\mu$ given Condition-\ref{assumption:Boltzmann 2}. If $\mu(\Gamma_\text{eq})$ gives the relative likelihood of finding the system in $\mu(\Gamma_\text{eq})$ then for all practical purposes $\Gamma_\text{eq}$ is a steady or \emph{equilibrium} state of the system because the system will almost always be found there. More significantly, if an equilibrium state exists, then a system that starts in a small macro-state will typically spend most of its future time in $\Gamma_\text{eq}$. The basic picture is depicted in Fig~\ref{fig:phase space}. This picture is plausible because the counting suggested by the required interpretation of $\mu$ immediately suggests that a system starting outside of $\Gamma_\text{eq}$ has little option but to quickly wander into $\Gamma_\text{eq}$, where it will remain for a very long time. But now there is a puzzle. Applying the same reasoning backwards in time suggests that a state finding itself in a small macro-state will also typically spend of all its \emph{past} in equilibrium. Because this apparently violates our knowledge that the past entropy of the universe was low, we are faced with the so-called \emph{second problem} of Boltzmann (see \cite{brown2001origins}). To solve this problem, one can posit an extremely \emph{atypical} condition on the earliest relevant state of the system. Under this condition, the system will typically find that it will approach the equilibrium state in the future. Note the temporal significance of the measure (Condition-\ref{assumption:Boltzmann 2}) and its central role in grounding the explanation of time asymmetry.

\subsection{The Past Hypothesis} 
\label{sub:the_past_hypothesis}

The main idea behind the PH is to evoke the Boltzmann-style reasoning of the previous section to explain time asymmetry in the universe. The system in question is then taken to be the entire universe and the PH itself translates into a special condition on the earliest relevant state of the universe. All of the mathematical quantities discussed above --- phase spaces, measures, macro-states, etc --- are then taken to represent aspects of the universe as a whole. The proposed explanation is given in terms of a typicality argument: universes that obey the appropriate PH, it is claimed, will typically evolve towards an equilibrium state in the future. Time-asymmetry arises by asymmetrically applying the special condition to past, rather than future, states. That the Boltzmann reasoning, whose empirical success is traditionally realised in closed sub-systems of the universe, can provide explanatory leverage when applied to the universe \emph{as a whole} is then taken as a further condition (Condition-\ref{assumption:typicality}) for the PH. Empirical support for the extreme atypicality of the initial state of our universe is taken to be implied by abundant cosmological evidence for a low-entropy early universe (e.g., the near-thermality of the Cosmic Microwave Background (CMB) power spectrum). We take this to be a final condition (Condition-\ref{assumption:observations}) for the viability of the PH.

\subsection{Requirements of the Past Hypothesis} 
\label{sub:key_assumptions_of_the_past_hypothesis}

We will now state all conditions identified in \S\ref{sec:preliminaries} (this list of conditions is \emph{not} intended to be sufficient for the PH).
\begin{enumerate}[(A)]
	\item There exists a measure, $\mu_\text{universe}$, on the phase space of the universe, $\Gamma_\text{universe}$, that is simultaneously:\label{assumption:measure}
	\begin{enumerate}[(\ref*{assumption:measure}1)]
		\item mathematically precise, \label{assumption:rigor}
		\item empirically unambiguous, and \label{assumption:uniqueness}
		\item invariant under all gauge symmetries. \label{assumption:invariance}
	\end{enumerate}
	\item It is justifiable to interpret the weights given by the chosen measure in terms of the relative likelihood of the system being in a given region at a given time.\label{assumption:Boltzmann 2}
	\item There is an epistemologically meaningful and mathematically well-defined projection from the microscopic phase space of the universe, $\Gamma_\text{universe}$, to a macroscopic phase space, $M_\text{universe}$.\label{assumption:epistomology}
	\item There exists a unique and exceptionally large state, defined to be the \emph{equilibrium state} $\Gamma_\text{eq}$, that is a typical macro-state on the phase space of the universe at any given energy $E$; i.e.,
		$$\frac{\mu_\text{universe}[\Gamma_\text{E,universe}] - \mu_\text{universe}[\Gamma_\text{eq}]}{\mu_\text{universe}[\Gamma_\text{E,universe}]} \ll 1\,.$$\label{assumption:Boltzmann 1}
	\item Typicality arguments have explanatory power when applied to the universe.\label{assumption:typicality}
	\item There is cosmological evidence for the PH being true.\label{assumption:observations} 
\end{enumerate}


\section{Criticisms of the PH} 
\label{sec:deconstructing_the_argument}

In this section we will set the stage for the arguments motivating the considerations of \S\ref{sec:symmetries_and_measure_ambiguities}. We identify and describe three categories of criticisms of the PH:
\begin{enumerate}[(I)]
	\item \emph{Mathematical precision}. These criticisms question whether the formal quantities necessary for stating the PH can be given precise, unambiguous mathematical definitions.\label{worry: ambiguity}
	\item \emph{Dynamical considerations}. These criticisms grant \eqref{worry: ambiguity} but question whether the resulting formal quantities have the physical characteristics required for a Boltzmannian explanation  --- especially when gravitational interactions are taken into account.\label{worry:intuition}
	\item \emph{Justification and explanation}. These criticisms grant both \eqref{worry: ambiguity} and \eqref{worry:intuition} but question the explanatory power and physical justification of the typicality arguments used when applied to the universe as a whole.\label{worry:typicality}
\end{enumerate}
Division of criticism into the above categories emphasizes the reliance of the latter forms of criticism on being able to provide adequate responses to the former. If, for example, one cannot meet the standards of Category-\ref{worry: ambiguity}, then the framework must be rejected and the considerations of Categories \ref{worry:intuition} and \ref{worry:typicality} become irrelevant. We will see below that there are already significant worries raised at the level of Categories \ref{worry: ambiguity} and \ref{worry:intuition} even though a significant amount of philosophical literature is focused on evaluating criticism falling into Category-\ref{worry:typicality}. We now discuss several examples, taken to be exemplars, of criticism to illustrate each of the above categories. This analysis will help illustrate the importance of the distinct properties of the Liouville measure that provided the basis for the dilemma presented in \S\ref{sub:symmetry_and_ambiguity}.

\subsection{Category I: mathematical precision} 
\label{sub:case_c_measure_ambiguities}

In this section we will primarily be concerned with issues arising from Conditions-\ref{assumption:measure} due to infinite phase spaces. Such phase spaces entail serious mathematical problems for measure-theoretic approaches to explanation. These problems stem from two distinct sources. The first arises because measures evaluated on an infinite interval can only be defined according to a limiting procedure that typically leads to physically significant regularization ambiguities. These problems are compounded in field theories because of a second source of ambiguity due to the phase space itself being infinite dimensional. In this case, it is a theorem that no Borel measure exists \citep{Curiel:2015oea} so that the system must be truncated to a finite phase space in order to accommodate any measure. Ambiguities of these two kinds lead to a tension between mathematical precision (Condition-\ref{assumption:rigor}) and empirical uniqueness (Condition-\ref{assumption:uniqueness}). To make matters worse, the purely mathematical problem of defining any measure on the phase space of general relativity invariant under all space-time symmetries is far from being solved. This open technical problem is in fact one of the main formal obstructions to obtaining a canonical formulation of quantum gravity. With this in mind, it is advisable to explore various approximations to general relativity that render the computations of measures more tractable. But even in this simplified setting, one encounters immediate and troubling difficulties that are emblematic of the more general case.

Pioneering work in \cite{gibbons1987natural} that was elaborated on by several authors in both the physics \citep{Hawking:1987bi,Hollands:2002xi,Corichi:2010zp,Ashtekar:2011rm,Wald:2012zf} and philosophy literature \citep{earman2006past,frigg2009typicality,Curiel:2015oea} shows that the natural measure on homogeneous and isotropic cosmologies has infinite phase space volume. In the references listed, different schemes are provided for handling these divergences, and these schemes introduce ambiguities. A particular illustration of this will be outlined in detail in \S\ref{sub:dynamical_similarity_in_the_universe}. To resolve these mathematical ambiguities (of the first kind discussed above), new inputs, which are often physical in nature, must be introduced. It is thus paramount that the extra inputs needed to resolve these ambiguities neither conflict with other symmetry principles, in accordance with Condition-\ref{assumption:invariance}, nor implicitly assume what is trying to be explained: i.e., the time-asymmetry of local thermodynamic processes. Otherwise, the explanatory power of the PH is undermined.

To illustrate the extent to which these ambiguities are problematic, consider the concrete results of different authors with different intuitions performing computations of the relative likelihood of cosmic inflation. Advocates for inflation \citep{Kofman:2002cj,Carroll:2010aj} proposed a measure according to which the probability of inflation was found to be infinitesimally close to 1. Inflation skeptics \citep{Turok:2006pa} proposed an alternative measure where the probability of inflation was found to be 1 part in $10^{85}$! This remarkably huge discrepancy reflects the extent to which individual beliefs can affect cosmologist's determinations of the appropriate physical principles used to justify their measure and the difficulties of resolving the tensions between Condition-\ref{assumption:rigor} and Condition-\ref{assumption:uniqueness}. Any conclusions drawn on the basis of a typicality argument must be assessed in light of such remarkable disagreement between cosmologists.

Ambiguities of this kind are not improved when more realistic models including cosmological inhomogeneities are considered. Any preliminary hopes, such as those alluded to in \cite{callender2010past}, that adding an infinite number of degrees of freedom would help resolve these ambiguities can be seen to be in vain when explicit models are considered. This has been done, for example, in \cite{Wald:2012zf}. What was found there was that the additional degrees of freedom introduce corresponding regularization ambiguities of the second kind discussed above. It is therefore necessary to introduce new physical principles in order to resolve these ambiguities. Given the daunting nature of a full general relativistic treatment, these considerations raises serious doubts regarding the possibility of being able to attribute any meaningful notion of typicality to the universe.


\subsection{Category II: dynamical considerations} 
\label{sub:case_b_gravitational_considerations}


In this section we will consider the unique properties of gravitational dynamics that complicate our entropic intuitions for the universe, assuming that a well-defined truncation of the phase-space exists on which a Liouville measure can be defined. Consider the equilibrium state of a free gas. It is smooth, homogeneous and nothing like the current state of the universe, which is characteristically clumpy and uneven. Those clumps comprise, amongst other things, star systems --- one of which supports the far-from-equilibrium biological system we find ourselves in. On the other hand, analysis of CMB temperature fluctuations reveals only a small $10^{-5}$ deviation from homogeneity. How can these observations be compatible with a low entropy past state? The standard response to this is that the gravitational contribution to the entropy should dominate at late times because of the unusual thermodynamic character of the gravitational interactions. This contribution is so great that it more than compensates for the decrease in entropy observed through the clumping of matter. Intuition for this comes from entropic considerations in Newtonian $N$-body self-gravitating systems, which have been used to model, for example, the dynamics of dust and stars in galaxies and galaxy clusters. But even in this simplified and well-tested setting there are difficulties that are emblematic of the considerations of \S\ref{sub:case_c_measure_ambiguities}.

Because Liouville volume is a volume on phase space, the inverse square potential due to gravity and the large momenta it can generate flip expectations for what constitutes a high and low entropy state. The steep gravitational potential well taps a vast reservoir of entropy allowing for the kind of sizable low entropy fluctuations we see in biological systems on Earth. These features as well as the difficulties they entail are reviewed nicely in \cite{Padmanabhan:2008overview,PADMANABHAN:1990book}, which gives detailed proofs of many of the results referenced below. This flipping of expectations is argued to occur not only for $N$-body systems, but also in a full-fledged general relativistic treatment of entropy. Thus, advocates of the PH (for example \cite{goldstein2004boltzmann,albert2009time}) emphasize the $N$-body intuition pump as providing an explanation for why the early homogeneous state of the CMB should be thought of as having low entropy and the current clumped state, which contains steadily accumulating stable records, as having high entropy. Moreover, this intuition was a primary motivation for early attempts at formulating an explicit PH such as Penrose's \emph{Weyl Curvature Hypothesis} \citeyearpar{Penrose:1979WCH}.

The $N$-body intuition pump, however, also raises potential concerns. Firstly, if we follow the past state far enough into the early universe, a full general relativistic treatment becomes unavoidable. But as we have already seen in \S\ref{sub:case_c_measure_ambiguities}, such a treatment suffers from troubling ambiguities and it is not clear that the simple Newtonian intuition will remain valid. Another significant worry is the definition of equilibrium itself. The notion of equilibrium in gravitational systems is complicated by two sources of divergence (for details see \cite{Padmanabhan:2008overview}): i) the infinite forces particles exert upon each other when they collide, and ii) the infinite distances particles can obtain when ejected from a system. To cure these divergences, it is necessary to render the entropy finite by imposing additional constraints. This involves closing the system at some maximum size, so that particles are not allowed to escape, and forbidding two particles from being able to collide. This requires extra assumptions that must be grounded in physically acceptable principles. It is therefore paramount that these physical idealizations be well-motivated. But the fact that these idealizations break down under specified conditions implies difficulties in defining stable equilibrium for the system. Indeed, $N$-body systems are known to only have local --- but no global --- maxima \citep{Padmanabhan:2008overview}. Thus, gravitating systems do not have genuine equilibrium states, and Condition-\ref{assumption:Boltzmann 1} cannot be strictly satisfied. In absence of an equilibrium state, thermodynamic quantities such as macro-states and their entropy cannot be defined and Condition-\ref{assumption:epistomology} is strictly violated. While this is not problematic for local meta-stable systems like a galaxy, it can certainly be problematic for globally defined systems like the entire universe. Moreover, even when local equilibria exist, there is still no guarantee that gravitational dynamics will actually steer the system towards these local equilibria in order to satisfy Condition-\ref{assumption:Boltzmann 2}. The crucial role of dynamics in the Boltzmannian argument has been emphasized in \cite{frigg2009typicality} and \cite{brown2001origins}.


\subsection{Category III: justification and explanation} 
\label{sub:case_a_typicality}

This section will firstly be concerned with the essential need to satisfy Condition-\ref{assumption:Boltzmann 2} by finding a valid justification for using Liouville volume as a typicality measure, assuming all concerns of Category~\ref{worry: ambiguity} and \ref{worry:intuition} have been resolved. In conventional statistical mechanical systems, this justification proceeds along two traditional routes. The first and oldest route relies on a theorem by \cite{birkhoff1931proof} that states that for ergodic systems the average time spent in a particular phase space region becomes roughly proportional to its Liouville volume if the timescales in question are much longer than the Poincar\'e recurrence time. Unfortunately, for almost all system --- and certainly for the universe --- the Poincar\'e recurrence time is significantly longer than the estimated time since the Big Bang. The second route, usually favored for its practicality, is to argue that the system undergoes a process called \emph{mixing}. Roughly speaking, a system is mixed when the long-run evolution of the measure of a system becomes approximately homogeneous, and therefore Liouvillian. Many systems exhibit this property and the relevant mixing timescales can be computed explicitly. Unfortunately, \cite{Wald:2012zf} argue that the observed expansion of the universe is too rapid to allow the large scale structures of the universe to interact often enough for mixing to occur on these scales. This suggests that it is unreasonable to expect the universe as a whole to undergo mixing. It would seem that in terms of conventional justification schemes for the Liouville measure Condition-\ref{assumption:Boltzmann 2} cannot be made compatible with the observational requirements of Condition-\ref{assumption:observations}.

It is possible to look for justification schemes satisfying Condition-\ref{assumption:Boltzmann 2} that do not originate from conventional statistical mechanical considerations. One proposal made by Penrose \citeyearpar{Penrose:1979WCH,penrose1994second} and later advocated  (either implicitly or explicitly) by \cite{goldstein2001boltzmann}, \cite{lebowitz1993boltzmann}, and \cite{albert2009time} is a version of the Principle of Insufficient Reason (PIR) as formalized by Laplace. In Penrose's version, a blind Creator must choose initial conditions for the universe among the space of all possibilities. Being indifferent to which conditions to choose, the Creator assigns equal likelihood to each possibility according to the Liouville measure. Given the failure of standard justifications schemes, \cite{Wald:2012zf} point to Penrose's proposal as the only available alternative. Unfortunately, the PIR has a troubled history in the philosophy of science and suffers from several well-known difficulties. At least four prominent criticisms are identified in \cite{uffink1995entropyconsistency}. While some of these are addressed implicitly throughout this text, one line of criticism dating back to Bernoulli is noteworthy because it also directly puts into question the validity of Condition-\ref{assumption:epistomology}. In this line of criticism one derives paradoxes that originate in an incompatibility between the measures obtained when applying the PIR to different choices of partition for the micro-states of a system. These paradoxes occur when the partitions correspond to \emph{disjunct coarse-grainings} or \emph{refinements} of each other \citep{norton2008ignorance}. There is nothing in the PIR that tells you which partitioning of the micro-states is the ``correct'' one precisely because this would require some non-trivial knowledge about how these partitions may have been gerrymandered. Without direct knowledge of the ``correct'' partitioning of micro-states, the PIR loses all explanatory power.


The only remaining justification for the Liouville measure is a uniqueness argument under time-symmetry. If one requires a time-symmetric measure, then the uniqueness of the Liouville measure under the requirement of being preserved by arbitrary Hamiltonian evolution does single it out. However, as we will see in \S\ref{sec:symmetries_and_measure_ambiguities}, very general symmetry considerations will put into doubt any motivations for using the Liouville measure to establish a notion of typicality for models in the universe.

We end this section by mentioning a prominent dialectic between Price \citeyearpar{price2002boltzmann,price2004origins} and Callender \citeyearpar{callender2004measures,callender2004origins} on the explanatory power of the PH that questions the validity of Condition-\ref{assumption:typicality}. In this dialectic Price argues that the PH itself should require explanation in pain of applying a ``temporal double standard'' to a past state when an atypical future state would plainly require explanation. Callender responds by stating that contingencies rarely (or never) require explanation, and an initial condition such as a PH is a contingency of this kind. 

\section{A Dilemma for the Past Hypothesis} 
\label{sec:symmetries_and_measure_ambiguities}

\subsection{Preliminaries: dynamical similarity as a gauge symmetry of the universe} 
\label{sub:dynamical_similarity_in_the_universe}

Before establishing the horns of the dilemma, it will be convenient to state some results that will be central to the analysis. We will need to give the definition of a particular symmetry of the universe and list some of its core properties. The symmetry that will be central to our argument is called \emph{dynamical similarity}. The three aspects of dynamical similarity that will be needed for our analysis are: first that dynamical similarity is a gauge symmetry of any general relativistic formulation of the laws of the universe, second that the Liouville measure is not invariant under dynamical similarity and third that in known theories of the universe dynamically similar measures are badly time-asymmetric. To illustrate our first point, we must show that dynamical similarity relates empirically indistinguishable descriptions of a general relativistic system. We will do this first by making a general argument and then by showing that this general argument is consistent with the treatment of particular cosmological theories.

We begin by giving a definition of dynamical similarity.\footnote{ For an excellent account of dynamical similarity and its role in defining measures in cosmology see \cite{Sloan:2018lim}. } Consider any system whose dynamical possibilities are specified by Hamilton's principle. For such systems, an action functional $S[\gamma]$ is given such that the Dynamically Possible Models (DPMs), $\gamma_\text{DPM}$, of the system are stationary points of $S$: 
\begin{equation}\label{eq:stationarity cond}
	\delta S[\gamma]|_{\gamma_\text{DPM}} = 0 \,.
\end{equation}
Then any transformation on the state space of such a system that rescales the action functional,
\begin{equation}\label{eq:dyn sym def}
	S \to c S\,,
\end{equation}
is defined to be a \emph{dynamical similarity}. For any system of this kind, a dynamical similarity will map a DPM to another DPM, and is therefore a symmetry. This follows straightforwardly from the fact that the stationarity condition \eqref{eq:stationarity cond} is invariant under \eqref{eq:dyn sym def}. Dynamical similarities are therefore symmetries of any general relativistic description of the universe because general relativity can be formulated in terms of Hamilton's principle.

This notion of symmetry, namely a transformation that maps DPMs to DPMs, is not yet enough for our argument. We will further need to show that dynamically similar DPMs are empirically indistinguishable. To see that this is true, observe that the constant in the transformation \eqref{eq:dyn sym def} can always be set to 1 by a suitable choice of units for the action. Since the unit of action is the unit of angular momentum, we find that dynamical similarities map DPMs to DPMs with different choices of units of angular momentum. Only if these choices can be compared with an external reference scale for angular momentum can the DPMs in question be empirically distinguished. If instead the units of angular momentum are referenced from within the system, then an arbitrary choice of units can have no empirical consequences. Because we are interested in a general relativistic description of the entire universe, there can be no external reference unit to distinguish between dynamically similar descriptions of the system. Thus, dynamical similarities are symmetries of a general relativistic description of the universe that relate empirically indistinguishable models; i.e., they are gauge symmetries.

This point is well-appreciated by cosmologists. In writing down the equations of cosmological systems, one starts with a general relativistic formulation and then imposes spatial homogeneity and isotropy. The simplest models of inflation can thus be described by a single geometric variable $v(t)$ representing the volume of a co-moving patch of the universe and a single massive scale field $\phi(t)$. The Hamiltonian for this system can be written as:
\begin{equation}
	\mathbbm H = \lf[  - H^2 + \frac{\pi_\phi^2}{v^2} + \tilde m^2 \phi^2 \rt]\,,
\end{equation}
where $H$ is \emph{Hubble} red-shift parameter conjugate of $v$, $\pi_\phi$ is the momentum of the scalar field, and $\tilde m$ is a dimensionless mass.\footnote{ To obtain this expression we have absorbed all units of angular momentum into the variables $v$, $\pi_\phi$ and $t$. Thus, $H$ and $\phi$ are dimensionless. We have also used a time parameter $t = v \tau$, where $\tau$ is the proper time along a homogeneous slice.}

This theory inherits a dynamical similarity from its underlying general relativistic description. If we remember that $S = \int \de t\, \lf( \dot v H + \dot \phi \pi_\phi - \mathbbm H \rt)$ then the transformation
\begin{align}\label{eq:dyn sim explicit}
	v &\to c v & \phi &\to \phi \\
	H &\to H & \pi_\phi &\to c \pi_\phi \notag\,,
\end{align}
is a dynamical similarity when $t \to ct$. The physical significance of the dynamical similarity \eqref{eq:dyn sim explicit} is straightforward to understand. It represents the freedom to arbitrarily choose the initial volume of a fixed fiducial cell while keeping the red-shift fixed. In cosmology, dynamical similarity therefore reflects the well-known property that the scale factor is an unobservable degree of freedom even though its momentum, the Hubble parameter, is observable. This achieves our first objective.

Our second objective is to show that the Liouville measure is not invariant under dynamical similarity. This together with the previous result will be essential for establishing the second horn of the dilemma: the breaking of gauge invariance by the Liouville measure. This can be achieved by expanding upon the mismatch between the transformation properties of the volume $v$ and its conjugate momentum $H$. The Liouville measure is a homogeneous measure on phase space. This means that it gives the same weight to a configuration variable as it does to the corresponding momentum. It is thus impossible for any of measure of this kind to be invariant under a symmetry that acts in an unbalanced way on the phase space variables. We can illustrate this explicitly for the cosmological theory given above. A set of canonically conjugate variables for this theory is: $\{v, H, \phi, \pi_\phi \}$, and therefore the Liouville measure is
\begin{equation}
	\mu_L(R) = \int_R \de v\, \de H\, \de \phi \, \de \pi_\phi\,.
\end{equation}
This measure is explicitly not invariant under the symmetry \eqref{eq:dyn sim explicit}. While illustrative and physically relevant, the non-invariance of the Liouville measure in this example is not just a special feature of this particular cosmological theory, but a general property of the Liouville measure. In order for a dynamical similarity to rescale the action as in \eqref{eq:dyn sym def} it must rescale the symplectic potential $\theta = p \de q \to c \theta$. But since the Liouville measure is just a power of the exterior derivative of the symplectic potential, $\mu_L(R) = \int_R \lf( \de \theta \rt)^n $, the Liouville measure itself will necessarily rescale under a dynamical similarity. Thus, the Liouville measure in general cannot be invariant under dynamical similarity.

The last objective of this section is to show that the lack of invariance of the Liouville measure results in a significant numerical time-asymmetry in its projection onto the dynamically similar state space relevant to cosmological theories. This result will be useful in strengthening the case for the loss of explanatory power that leads to the first horn of the dilemma (see \S\ref{sub:the_origin_of_measure_ambiguities_in_cosmology} for details). To achieve the last objective, we will recall the results of well-known derivations.\footnote{For a summary of the results used here see \cite{Wald:2012zf}.} The measure that is relevant to our considerations is a measure not on the space of states but on the space of models. This can be achieved by projecting the Liouville measure onto some initial data surface on phase space. Because the Liouville measure is time-independent, the choice of initial data surface is arbitrary. For the cosmological theory presented in this section, a convenient choice of initial data surface that is also empirically meaningful is that of a surface of constant red-shift: $H = H^\star$. This choice leads to the Gibbons--Hawking--Stewart measure \citep{gibbons1987natural}
\begin{equation}\label{eq:GHS}
	\mu_\text{GHS}(r) = \int_r \sqrt{ (H^\star)^2 - \tilde m^2 \phi^2 } \de v\, \de \phi\,,
\end{equation}
where $r$ is a region on the surface $H = H^\star$ that is compact in $\phi$ but not in $v$. This measure is not regarded to be physical in part because of its non-compact domain in terms of $v$ but, more importantly, because of the arbitrariness of the value of $v$ in terms of a choice of initial fiducial cell. More recently, \cite{Sloan:2019wrz} has established a direct link between this arbitrariness and dynamical similarity.\footnote{ The connection was first noticed in the context of Loop Quantum Cosmology by \cite{Corichi:2010zp} and \cite{Ashtekar:2011rm}. } To obtain a physically significant measure, \cite{Hawking:1987bi} defined a regularization procedure that takes advantage of the homogeneity of \eqref{eq:GHS} in $v$ to integrate over all possible values of $v$. The resulting measure
\begin{equation}\label{eq:prob inflation}
	\text{Prob}(r_\phi) = \lim_{v_\text{max} \to \infty} \frac{ \int_0^{v_\text{max}} \de v }{  \int_0^{v_\text{max}} \de v  } \frac{  \int_{r_\phi} \de \phi \sqrt{ (H^\star)^2 - \tilde m^2 \phi^2 }}{ \int_{r_{\phi_\text{max}}} \de \phi \sqrt{ (H^\star)^2 - \tilde m^2 \phi^2 }} \to \text{finite} 
\end{equation}
is finite. The result depends only on the ratio of the integrals over the region $r_\phi$, which can be used to define inflation, and the finite region $r_{\phi_{\text{max}}}$, which is given in terms of the dynamical constraints of the theory. From the perspective of dynamical similarity, the integration over $v$ is motivated by requiring that the physical measure be invariant under symmetries that relate physically indistinguishable models. The integral over $v$ is an integration over the action of the dynamical similarity \eqref{eq:dyn sim explicit}. The physical measure \eqref{eq:prob inflation} is therefore invariant under \eqref{eq:dyn sim explicit} while the unphysical measure \eqref{eq:GHS} is not.

The integration over $v$ creates a new problem. The physical measure \eqref{eq:prob inflation} depends explicitly on the choice of initial data surface as determined by the choice of initial red-shift factor $H^\star$. This dependence on $H^\star$ is significant. As was shown explicitly in \cite{Wald:2012zf}, the different choices of $H^\star$ used by inflation sceptics \citep{Turok:2006pa} compared with inflations advocates \citep{Kofman:2002cj,Carroll:2010aj} leads to a colossal $85$ order of magnitude difference between the estimates of the likelihood of inflation. Because a choice of $H^\star$ corresponds to a choice of initial time, this huge numerical imbalance leads to a significant temporal asymmetry: choosing a more recent value of $H^\star$ gives a dramatically smaller value for the weight of the same region $r_\phi$.

This result is not just a special feature of the particular cosmological theory developed in this section. The Liouville measure is the unique time-independent measure on phase space. But, as we have shown, the Liouville measure is in general not invariant under dynamical similarity. There is therefore no (smooth) time-independent measure invariant under dynamical similarity. This means, in general, that a dynamically similar measure on the space of models will necessary depend on the choice of initial data surface (e.g., it will depend on $H^\star$). Moreover, the temporal asymmetry introduced by this is significant. For the theory introduced in this section, it leads to an $85$ order of magnitude difference between different choices of $H^\star$. There are good reasons to believe that this numerical imbalance will persist in any general relativitistic description of the universe. The interpretation of dynamical similarity in terms of an arbitrary choice of volume will persist in general relativity. In this context, the red-shift factor $H$ is still the variable conjugate to $v$. The temporal asymmetry will then always depend on the initial choice of $H^\star$, and this varies wildly between now and the empirically accessible past in a monotonic way. The huge monotonic variation of the Hubble parameter over the known history of the universe therefore introduces a significant time asymmetry into the definition of a dynamically similar measure.

\subsection{The first horn: loss of explanatory power} 
\label{sub:the_origin_of_measure_ambiguities_in_cosmology}

The analysis of \S\ref{sec:deconstructing_the_argument} has established that there are many concerns regarding the justification of the choice of typicality measure used to formulate a PH. In \S\ref{sub:case_b_gravitational_considerations} it was argued that self-gravitating systems have unusual thermodynamic properties and in \S\ref{sub:case_a_typicality} these arguments where combined with known facts about the universe to suggest that conventional statistical mechanical justifications fail when applied to the universe. Justifications that rely on indifference principles where also criticised on epistemological grounds. The analysis of \S\ref{sec:deconstructing_the_argument} therefore leads to the conclusion that the only tenable justification for choosing the Liouville measure is an argument from time-independence. The Liouville measure is indeed singled out as being the unique measure on phase space that is preserved by an arbitrary choice of dynamics. At first sight this uniqueness appears to be particularly convenient because a time-independent measure is very natural in the context of a PH. But time-independence in the measure is more than a question of convenience in the context of a PH. In fact, it is an essential ingredient for the PH independent of any other justificatory considerations.

Following \cite{price2002boltzmann}, the logic of the PH presented in \S\ref{sec:preliminaries} constitutes a contrastive explanation of the form: if A then B rather than C. The explanans A --- i.e., the PH itself --- is taken to explain the explanandum B --- i.e., the fact that typical processes are seen to overwhelmingly occur in a time-asymmetric way. The outcome C is then a typical member of a contrast class of outcomes that would be likely if not for A. The explanatory power of A comes from increasing the likelihood of B relative to C. In the case of a PH, the contrast class is the set of worlds where typical processes overwhelmingly occur in a time-symmetric way. According to this logic, in order for the PH to be a good explanation of time-asymmetry, it must be the only significant source of time-asymmetry. Clearly this is consistent with the apparent time-symmetry of the form of the fundamental laws. This consistency however is not sufficient. When a time-asymmetric measure is introduced into the formalism, the time-asymmetry of the measure could itself provide an explanation for the time-asymmetry of typical processes. This is especially true if the time-asymmetry of the measure introduces a significant numerical temporal gradient as was shown in the previous section for the case of cosmological models. Moreover, the time-dependence of the measure introduces an ambiguity in terms of which instant should be used in order to obtain a measure on the space of models. Such an ambiguity can only be resolved by including some additional principle to the PH --- thus undermining much of its explanatory appeal. It is therefore essential to the logic of the PH that the measure employed be time-independent, and especially important that the measure not be badly time-asymmetric. Otherwise we would have no reason to believe that processes would not occur in a time-asymmetric way even if the PH were not true. Note that these considerations hold regardless of any other justificatory considerations regarding the measure. This establishes the first horn of the dilemma.



\subsection{The second horn: violation of a gauge symmetry} 
\label{sub:symmetry_and_ambiguity}

In the preliminary \S\ref{sub:dynamical_similarity_in_the_universe} we saw that the projection of the Liouville measure onto the space of models, while time-independent, is nevertheless considered by cosmologists to be unphysical. Contrastingly, the measure that is considered by cosmologists to be physical was found to be invariant under dynamical similarity. We will now argue that this result is to be expected in any general relativistic description of the universe. To do this, we will show that a measure that is not invariant under symmetries that relate physically indistinguishable descriptions of a system (Condition-\ref{assumption:invariance}) introduces two distinct problems: first it introduces a distinction without difference and, second, it runs against standard practice in particle and statistic physics physics.

Consider a region $R$ that lives in the domain $\mathcal D(\mu)$ of some measure $\mu$ and a transformation $T: \mathcal D(\mu) \to \mathcal D(\mu)$ that maps this domain onto itself. Our assumptions demand that $T$ map states of a system to empirically indistinguishable states. The set of states in the region $R$ is therefore empirically indistinguishable from the set of states in the transformed region $R' = T(R)$. In general, the non-invariance of $\mu$ under $T$ implies that the weight of the transformed region is not necessarily equal to the weight of the original: $\mu(R) \neq \mu(R')$. But if this is true then the weights $\mu(R)$ and $\mu(R')$ provide a distinction at the representational level between the regions $R$ and $R'$. Given our original assumptions, this distinction cannot represent any empirical difference. In this sense, the measure $\mu$ therefore introduces a representational distinction that can't be captured by the empirical properties of the world. It is therefore \emph{not} a valid measure for describing empirical phenomena.

This argument is reinforced by standard practice in particle and statistical physics that requires that physical measures be invariant under all the gauge symmetries of a system. In the standard model of particle physics the gauge-invariance of the path-integral measure is a central foundational principle of the theory. More generally, the Faddeev--Popov determinant, which enforces the gauge-invariance of the path-integral measure, is considered a necessary ingredient in gauge theory (see \cite[Chap 15]{Weinberg:1996kr} for an overview and defence of this standard practice). Similarly in statistical physics, \cite{jaynes1973well} has argued influentially that measures should be invariant under transformations that relate indistinguishable states of a system. We therefore conclude that there are strong epistemological and methodological motivations for requiring Condition-\ref{assumption:invariance}.

We are now in a position to state the second horn of our dilemma. As we have shown in the previous section, dynamical similarity is a symmetry that maps states of any general relativistic description of the universe to indistinguishable states. Given the argument above, any measure not invariant under such a symmetry must violate a gauge symmetry and introduce a distinction without difference. Therefore, a measure on the state space of a generally relativistic description of the universe that is not dynamically similar will run into the symmetry-violating horn. But as was shown in \S\ref{sub:dynamical_similarity_in_the_universe}, the Liouville measure is not dynamically similar. It follows that use of the Liouville measure therefore violates a gauge symmetry of the theory. This is the second horn.

We now recall the first horn of the dilemma. The formulation of the PH must make use of the unique time-independent Liouville measure in order to retain its explanatory power. But the Liouville measure is not dynamically similar, and therefore introduces a distinction without difference. An advocate of the PH must therefore face the dilemma stated in the introduction: either lose explanatory power or introduce a distinction without difference.


\section{Discussions/Conclusions} 
\label{sec:prospectus}

We have seen that Boltzmann-style explanations of time-asymmetry that make use of a PH depend upon a series of very restrictive conditions. Our analysis in \S\ref{sec:deconstructing_the_argument} has uncovered several good reasons to question whether these conditions can ever be simultaneously satisfied. Broadly speaking we found that the nature of the phase space, dynamics and symmetries of general relativity provide reasons for pessimism regarding the prospects for providing and justifying a satisfactory notion of typicality for models of the universe. A common response against critiques of this kind is to observer that strict insistence on mathematical rigour has often been unreasonable in the development of theoretical physics. Controversy over difficult technical problems such as defining a measure on the solution space of general relativity should not, it is argued, halt progress altogether. It should still be reasonable to advance conjectures regarding the plausible features of measures that may one day become available.

While such a strategy --- effective or not --- is available in response to much of the analysis of \S\ref{sec:deconstructing_the_argument}, it is no longer available in response to the dilemma of \S\ref{sec:symmetries_and_measure_ambiguities}. This is because the dilemma is the result of a simple symmetry argument applied to a very general way of formulating the laws of the universe. To reject dynamical similarity is to reject a description of the physics of the universe in terms of Hamilton's principle. To reject the uniqueness arguments for the time-symmetry of Liouville's measure is to reject a description of the universe in terms of a phase space. To not require the gauge-invariance of the measure is to introduce a distinction without difference and to reject standard practice in particle and statistical physics. None of these escape routes is particularly appealing. Even if one grants all the technical assumptions required by the PH, the dilemma persists. On the other hand, a rejection of the PH as an explanation for time-asymmetry avoids the dilemma completely. But how then is one to explain the time-asymmetry of macroscopic processes given the apparent time-symmetry of the fundamental laws? In other words, how is one to solve the original problem of the arrow of time?

One possibility would be to embrace the necessary time-dependence of the measure implied by dynamically similarity. While the equations of motion of general relativity, and in particular the cosmological models discussed in \S\ref{sub:dynamical_similarity_in_the_universe}, are formally invariant under time-reversal, they also contain redundancy under dynamical similarity. The construction of a time-asymmetric measure invariant under dynamical similarity can be constructed for a very general class of systems \citep{Sloan:2018lim} in a way that mirrors the derivation of the physical measure \eqref{eq:prob inflation}. The resulting time-asymmetry of the measure can be shown to result from the non-conservative, time-irreversible structure of the reduced Hamiltonian for the system. Perhaps then the apparent time-symmetry of general relativity is simply an artefact of a representational redundancy? But if time-asymmetry really is built into the character of the empirically relevant formulation of the law, then this could provide a new basis for providing an explanation for the arrow of time. Such a strategy would parallel and further develop the approach suggested in \cite{Barbour:2014bga}, which also makes use of dynamical similarity. An important aspect of this approach is an account of the low-entropy past state as a generic, rather than highly atypical, feature of the theory. Such a scenario would therefore not require any PH. What remains is to extend a program of this kind to general relativity and to show that the time-asymmetry of the reduced system is indeed sufficient for explaining the observed time-asymmetry of macroscopic processes. This possibility opens up new and exciting directions for future investigations.




\section*{Acknowledgements}

I would like to thank Karim Th\'ebault for an enormous amount of encouragement, feedback, and helpful discussions. My thinking about the arrow of time has been heavily influenced by conservations with David Sloan, Tim Koslowski, Flavio Mercati and Julian Barbour. I'm also grateful to Roman Frigg, Fred Muller, Guido Bacciagaluppi, and audiences in Utrecht and Groningen for many useful discussions and feedback. Finally I'd like to thank Erik Curiel for valuable comments on an early version of the draft as well as Jan--Willem Romeijn and Simon Friederich for guidance, suggestions and mentorship. My work is supported by a Young Academy Groningen Scholarship.

\bibliographystyle{chicago}
\bibliography{philbib}

\end{document}